\documentclass{webofc}
\usepackage[varg]{txfonts}   
\begin{document}
\title{Galaxy clusters morphology with Zernike polynomials: the first application on \textit{Planck} Compton parameter maps}
%
%

\author{\lastname{V. Capalbo}\inst{1}\fnsep\thanks{\email{valentina.capalbo@uniroma1.it}} \and
        \lastname{M. De Petris}\inst{1} \and
        \lastname{W. Cui}\inst{2,3} \and
        \lastname{A. Ferragamo}\inst{1} \and
        \lastname{F. Ruppin}\inst{4} \and
        \lastname{G. Yepes}\inst{3}
}

\institute{Dipartimento di Fisica, Sapienza Università di Roma, Piazzale Aldo Moro 5, I-00185 Roma, Italy
\and
           Institute for Astronomy, University of Edinburgh, Royal Observatory, Edinburgh EH9 3HJ, UK
\and
           Departamento de Física Teórica and CIAFF, Módulo 8, Facultad de Ciencias, Universidad Autónoma de Madrid, E-28049 Madrid, Spain 
\and
           University of Lyon, UCB Lyon 1, CNRS/IN2P3, IP2I Lyon, France
           }

\abstract{%
  Insert your english abstract here.
}
\abstract{The study of the morphology of 2D projected maps of galaxy clusters is a suitable approach to infer, from real data, the dynamical state of those systems. We recently developed a new method to recover the morphological features in galaxy cluster maps which consists of an analytical modelling through the Zernike polynomials. After the first validation of this approach on a set of high-resolution mock maps of the Compton parameter, $y$, from hydrodynamically simulated galaxy clusters in {\sc The Three Hundred} project, we apply the Zernike modelling on $y$-maps of local ($z<0.1$) galaxy clusters observed by the \textit{Planck} satellite. With a single parameter collecting the main information of the Zernike modelling, we classify their morphology. A set of mock \textit{Planck}-like $y$-maps, generated from {\sc The Three Hundred} clusters, is also used to validate our indicator with a proper dynamical state classification. This approach allows us to test the efficiency of the Zernike morphological modelling in evaluating the dynamical population in the real \textit{Planck} sample.
}
\maketitle
\section{Introduction}
\label{intro}
It is well known that galaxy clusters are complex systems in continuous evolution and often far from conditions of thermodynamic equilibrium. Therefore, when some simplified assumptions such as hydrostatic equilibrium are adopted to estimate their masses, some biases arise between the real mass value and that derived from observations \citep[see e.g.][]{Gianfagna2021}. A study of the dynamical state is then crucial to classify galaxy clusters in order to obtain a reliable mass estimation \citep[see e.g.][]{Gianfagna2023}. 
A simple approach of the dynamical state evaluation is the analysis of maps of galaxy clusters generated at different wavelengths. 
We recently introduced a new method of morphological analysis which consists of modelling the cluster maps with a set of orthogonal functions, the Zernike polynomials (ZPs) \cite{Capalbo2021}. At first, we validated this approach on a set of clean mock maps of the thermal Sunyaev-Zel'dovich (SZ) effect \cite{SZ}, i.e. Compton-$y$ maps, realized for massive hydrodynamically simulated galaxy clusters ($M_{200} >6\times10^{14} h^{-1} M_{\odot}$) in {\sc The Three Hundred} (The300, hereafter) project \footnote{https://the300- project.org} \cite{Cui2018}.
We exploited the 3D information for these clusters to define the dynamical state \textit{a priori} and with a proper theoretical indicator, the $\chi$ parameter from \cite{Haggar2020,DeLuca2021}. Then, we applied the Zernike modelling on each $y$-map estimating a single parameter, $\mathcal{C}$, to quantify the morphology. All the details of this analysis are described in \cite{Capalbo2021}. We estimated a correlation of $\sim 60 \%$ between the dynamical indicator $\chi$ and the morphological parameter $\mathcal{C}$ computed on high-resolution ($5.25^{\prime\prime}$) $y$-maps for clusters at $z=0$. We also studied the variation of the correlation when moving to higher redshifts, up to $z=1$, and varying the angular resolution in the maps, up to 5 arcmin. 
Now we apply the method on a set of real $y$-maps for galaxy clusters in the \textit{Planck}-SZ sample and we compare the results with the analysis of mock \textit{Planck}-like maps realized for a sample of The300 clusters representative of the real sample and classified for the dynamical state with $\chi$, as described before. In this way, we estimate the reliability of our morphological analysis to infer clusters' dynamical states for \textit{Planck}.
 
\section{Zernike polynomials: definition and application}
\label{ZPs}
The ZPs are a complete and orthogonal set of functions defined on a unit disk. Their analytical expression is given by \cite{Noll1976}
\begin{equation}
    \label{eq:Zpol}
    Z^m_n(\rho,\theta) = N^m_n R^m_n(\rho) \cos{(m\theta)}\,\,\, , \,\,\,
    Z^{-m}_n(\rho,\theta) = N^m_n R^m_n(\rho) \sin{(m\theta)}
\end{equation}
where $\rho$ is the normalized radial distance ($0\leq\rho\leq1$), $\theta$ is the azimuthal angle ($0\leq\theta\leq2\pi$), $n$ and $m$ ($m\leq n$ and $n-m$\,=\, even) are the polynomial order and the angular frequency, respectively, $N^m_n=\sqrt{2(n+1)/(1+\delta_{m0})}$ is the normalization factor and $R^m_n(\rho)=\sum_{s=0}^{(n-m)/2} \frac{(-1)^s (n-s)!}{s!\Bigl(\frac{n+m}{2}-s\Bigr)!\Bigl(\frac{n-m}{2}-s\Bigr)!} \rho^{n-2s}$ is the radial term. 
Due to their properties, the ZPs can be efficiently used to fit, with a desired accuracy, arbitrary functions defined on circular domains. In particular, the orthogonality makes sure that there are no overlaps between the different terms of a set of ZPs. Here we use these functions to model Compton-$y$ maps for galaxy clusters. Then, $y(\rho,\theta)$ can be expressed as:
\begin{equation}
    \label{eq:fit}
    y(\rho,\theta)=\sum_{n=0}^N \sum_{m=0}^n c_{nm}Z_{nm}(\rho,\theta)
\end{equation}
where $N$ is the number of ZPs used in the modelling and $c_{nm}$ are the Zernike coefficients of the single polynomials.
A power spectrum analysis allows to chose the maximum order of the polynomials, $n$, to correctly model the requested spatial scales \cite{Svechnikov2015}. However, we want to highlight that the aim of this approach is not to exactly reconstruct the cluster maps, but rather to recognize the main features that can be related to different dynamical states, such as regular (mostly circular) patterns or inhomogeneities in the distribution of the signal in the maps, signs of possible relaxed or disturbed systems, respectively. In \cite{Capalbo2021} we analyzed high-resolution ($5.25^{\prime\prime}$ at $z=0$) mock $y$-maps, without noise contamination, therefore we preferred to use a large set of ZPs (45 terms, up to the order $n=8$) to model in detail the maps within a circular aperture of radius equal to $R_{500}$\footnote{$R_{500}$ is the radius within which the overdensity of the cluster is 500 times the critical density of the Universe at the cluster redshift.} for each cluster. For that model, we estimated a spatial resolution of $\sim 0.5 R_{500}$. Here, we use real maps, with residual noise and at lower resolution ($10^{\prime}$), in which the region inside $R_{500}$ is poorly resolved therefore, we reduce the number of ZPs in the fit. Following the criterion derived in \cite{Svechnikov2015}, if the number of ZPs decreases (increases) of a certain factor, the spatial resolution obtained in the modelling will increase (decrease) of the same factor. Based on this assumption, we reduce the order of ZPs used in this analysis by a factor of 4 with respect to \cite{Capalbo2021}, then we use 6 terms, up to the order $n=2$, estimating a spatial resolution of $\sim 2R_{500}$ that we verify is enough to appreciate different morphologies in the \textit{Planck} $y$-maps, remaining poor sensitive to the noise.
 
\section{Data sets}
\label{data}
In this Section, we describe the sample of galaxy clusters selected from the \textit{Planck}-SZ catalogue and from The300 project with their $y$-maps.

\begin{figure}[h]
\centering
\includegraphics[scale=0.5]{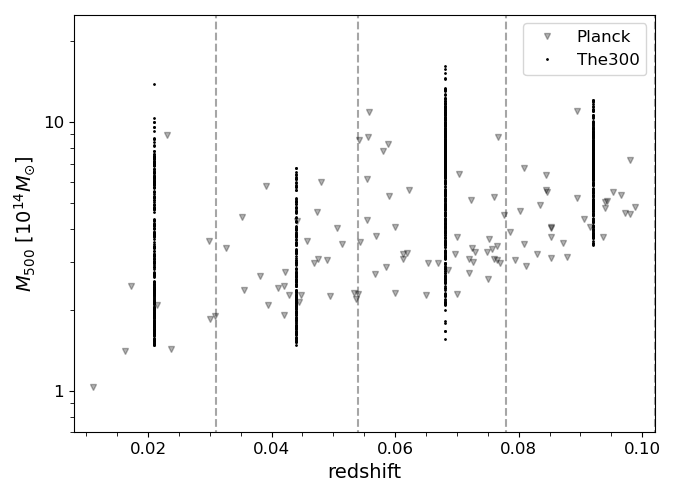}
\caption{\textit{$M_{500}-z$} distribution of \textit{Planck} (grey triangles) and The300 (black points) selected clusters. The vertical dashed lines in grey mark the redshift bins in which we divided the samples.}
\label{fig:sample}
\end{figure}

\begin{itemize}

    \item {\bf \textit{Planck}-SZ cosmology catalogue} The cosmology catalogue is composed of galaxy clusters detected by three different algorithms: two Matched Multi-Filters techniques, MMF1 and MMF3, and the PowellSnakes method, as described in \cite{PSZ2}. They are objects detected with S/N\,$\geq6$ in a patch of 65\% of the sky that excludes the galactic plane and point sources. The purity and completeness of the catalogue make it high-reliable for the cosmological parameter analysis \cite{Planck_n_counts}. From this catalogue, we select clusters at $z<0.1$ and with angular size $\theta_{500}=R_{500}/D_A(z)$ ($D_A(z)$ is the angular diameter distance) larger than the angular resolution of the $y$-maps, i.e. $10^{\prime}$ \cite{Planck_y_maps}. For each cluster, we extract gnomonic projections from the all-sky $y$-maps, centred on the cluster coordinates and with a circular aperture of radius equal to $R_{500}$. We use the two publicly available maps realized with MILCA \cite{Hurier2013} and NILC \cite{Remazeilles2011} component separation algorithms. Our final sample is composed of 109 galaxy clusters.
    Note that the $M_{500}$ reported in the catalogue for each cluster is derived by using the scaling relation $Y_{500}-M_{500}$ assuming a bias $(1-b)=0.8$ between the true mass and the hydrostatic mass \cite{Planck_bias}. We correct the mass values in the catalogue for the bias above, in order to obtain an estimate of the true mass of the clusters to be compared with the true mass of the simulated clusters (see figure~\ref{fig:sample}).
    
    \item {\bf Mock sample} For The300 galaxy clusters we realized mock $y$-maps with the PyMSZ code\footnote{https://github.com/weiguangcui/pymsz} \cite{Cui2018}. The maps are convolved with a $10^{\prime}$ FWHM and gridded in pixels of $1.7^{\prime}$, to mimic the real \textit{Planck} $y$-maps. In addition, a full-sky noise map is realized by using the \textit{Planck} noise power spectrum and the publicly available full-sky maps of the standard deviation of the Compton parameter, from which patches are extracted randomly to generate the final noise added in the $y$-maps. Note that the mock maps do not include point source contamination \citep[see][for details]{deAndres2022}.
    We analyze $y$-maps for The300 clusters selected in 4 snapshots of redshift ($z=0.02, 0.04, 0.07, 0.09$) to cover the redshift range of the \textit{Planck} clusters and we select only clusters with S/N $\in[\min$(S/N)$_{Planck}$, $\max$(S/N)$_{Planck}]$ in each snapshot.
    As for the real \textit{Planck} $y$-maps, we apply the Zernike fitting on the mock $y$-maps within a circular aperture of radius $R_{500}$. The \textit{Planck} and The300 cluster samples are shown in the $M_{500}-z$ plane in figure~\ref{fig:sample}.
\end{itemize}

\begin{figure}[h]
\centering
\includegraphics[scale=0.5]{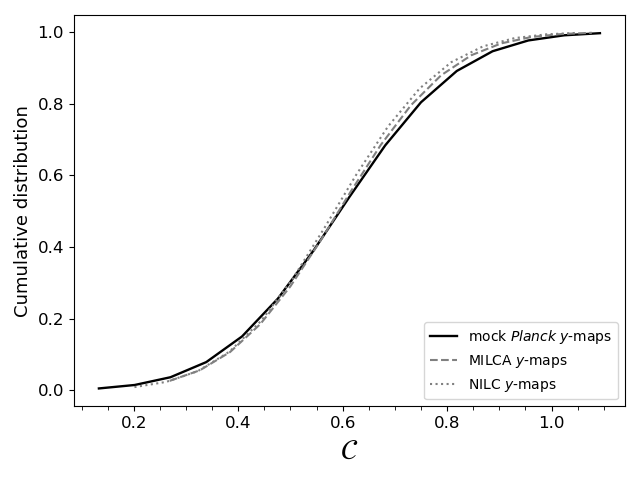}
\caption{Cumulative distribution of the $\mathcal{C}$ parameter from the Zernike fitting applied on mock \textit{Planck} $y$-maps (solid black line) for The300 clusters and MILCA (dashed grey line) and NILC $y$-maps (dotted grey line) for the real clusters.}
\label{fig:cumulative}
\end{figure}

\section{Results}
\label{results}
\begin{figure}[h]
\centering
\includegraphics[scale=0.49]{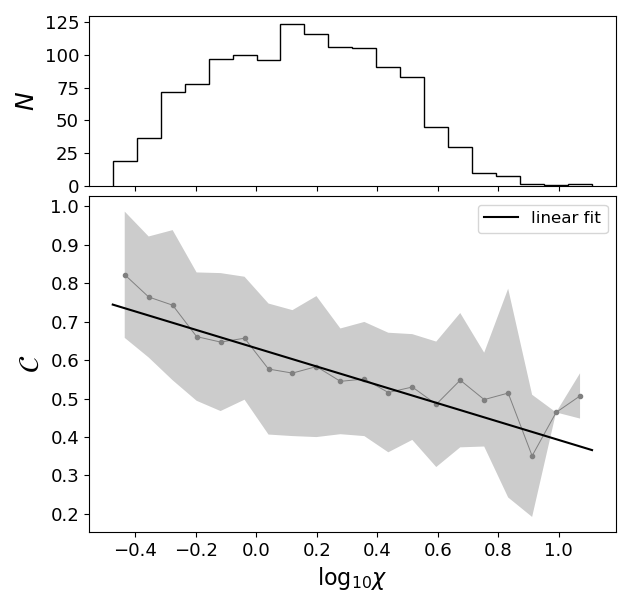}
\caption{\textit{Top panel}: Distribution of the number of The300 clusters selected, along $\log_{10}\chi$. \textit{Bottom panel}: $\mathcal{C}$ \textit{versus} $\log_{10}\chi$ for The300 clusters. The grey points are the mean values of the $\mathcal{C}$ parameter in each bin of $\log_{10}\chi$, while the grey area refers to $\pm1\sigma$. The solid black line is the best linear fit with the equation: $C=(-0.24\pm0.02)\log_{10}\chi + (0.63\pm0.01)$.}
\label{fig:fit}
\end{figure}

\begin{figure}[h]
\centering
\includegraphics[scale=0.5]{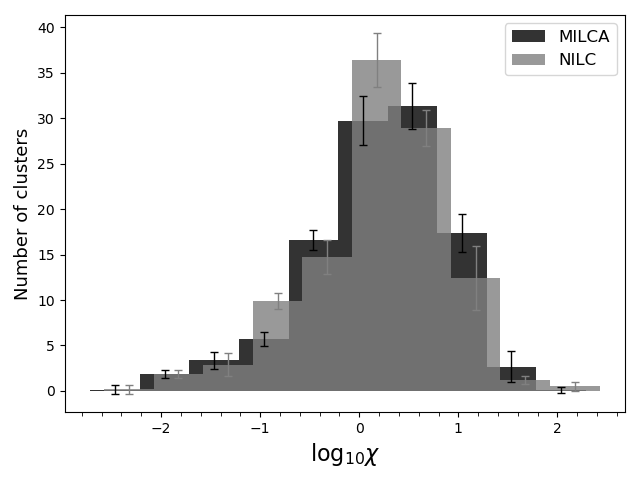}
\caption{Distribution of the number of clusters from the selected \textit{Planck}-SZ sample along $\log_{10}\chi$. The $\chi$ parameter is computed from the best linear fit, $\mathcal{C} = a \log_{10}\chi + b$, shown in figure~\ref{fig:fit}, by applying a Monte-Carlo error propagation for the $a$ and $b$ parameters. The height of each bar is the mean number of clusters in each bin of $\log_{10}\chi$ while the vertical lines on the top of each bar are at $\pm1\sigma$. The two colours, black and grey, refer to the distributions obtained from $\mathcal{C}$ computed on MILCA and NILC $y$-maps, respectively.}
\label{fig:chi_param}
\end{figure}

We fit the $y$-maps of the real and simulated clusters with the ZPs (here we use 6 polynomials, as described in Sect.~\ref{ZPs}) and we compute the $\mathcal{C}$ parameter as defined in \cite{Capalbo2021}: \begin{equation}
    \label{eq:C}
    \mathcal{C}=\sum_{n,m\neq0} |c_{nm}|^{1/2}
\end{equation}
In this definition the terms with $m=0$ are neglected because their contribution is almost invariant when modelling different maps in a wide range of morphologies, then they are not reliable for a morphological classification, as shown in \cite{Capalbo2021}. In figure~\ref{fig:cumulative} we show the cumulative distribution of the $\mathcal{C}$ parameter computed both on the mock and the real \textit{Planck} $y$-maps. The distributions are well-compatible, confirming that the mock maps selected from The300 sample are well representative of the real sample. 
The300 clusters are also classified for their dynamical state with a 3D indicator, the $\chi$ parameter, as defined in \cite{DeLuca2021}. This allows us to study the correlation, if any, between the Zernike morphological analysis of the 2D maps and the dynamical state classification available from the simulations.
We estimate a linear correlation of $38\%$ between $\mathcal{C}$ and $\log_{10}\chi$ for The300 clusters then, we compute the best linear fit
\begin{equation}
    \label{eq:linear_fit}
    \mathcal{C} = a \log_{10}\chi + b
\end{equation}
where $a=-0.24\pm0.02$ and $b=0.63\pm0.01$. The $\mathcal{C}$ parameter \textit{versus} $\log_{10}\chi$, binned in $\log_{10}\chi$, is shown in figure~\ref{fig:fit}, along with the best linear fit (solid black line).
The best linear fit is then used to convert the $\mathcal{C}$ distribution of the real \textit{Planck} clusters in a $\chi$ distribution, i.e. to convert the morphological analysis in a dynamical-state classification, as shown in figure~\ref{fig:chi_param}. Following the criterion in \cite{DeLuca2021} that relates $\log_{10}\chi>0$ to relaxed clusters, we estimate a fraction of $\sim65\%$ of relaxed clusters in the \textit{Planck} sample. 

To test the impact of the noise in the morphological analysis we also apply the Zernike fitting on another set of mock $y$-maps realized for The300 clusters already selected. These are maps only convolved with a $10^{\prime}$ beam but without adding noise realization. As expected, the $\mathcal{C}$ parameter estimated on these maps has a higher correlation ($47\%$) with the $\chi$ indicator. However, the best-fitting parameters $a$ and $b$ are compatible within $2\sigma$ with the values estimated when considering the proper \textit{Planck}-like maps with noise contamination. This means that the (low) number of ZPs used in the modelling allows it to be poor sensitivity to the noise and to remain however able to reveal the main morphological features for a dynamical classification.

\section{Conclusions}
\label{conclusions}
The study of the morphology of galaxy clusters maps is a reliable approach to estimate, from real data, their dynamical state. The morphological analysis that we developed employs a set of functions, the Zernike polynomials, that are well adapted for modelling cluster maps, due to their simple analytical properties. When applying the method to real data, the number of functions to use in the modelling must be tuned by considering the angular resolution in the maps and the noise contamination. The first application of this method on real data is done by using Compton parameter maps for galaxy clusters at $z<0.1$ observed by the \textit{Planck} satellite through the Sunyaev-Zel'dovich effect. 
To quantify the morphology of the cluster maps we compute a single parameter, $\mathcal{C}$, derived from the Zernike fitting.
We use a set of mock maps representative of the real sample, in terms of signal-to-noise and of $\mathcal{C}$ parameter, to validate the analysis. The mock maps, in fact, are realized for simulated clusters already classified for their dynamical state with a 3D indicator, $\chi$. We estimate a linear correlation of $38\%$ between $\mathcal{C}$ and $\log_{10}\chi$ for the simulated clusters. Then, we use this correlation to derive, from a linear fit, a $\chi$ distribution for the real clusters in the \textit{Planck} sample, from which we estimate a fraction of $\sim65\%$ of relaxed clusters. All the details of this application will be described in a paper in preparation.

%

\begin{thebibliography}{}
%
%
\bibitem{Gianfagna2021} 
Gianfagna G. \textit{et al.}, MNRAS, \textbf{502}, 5115 (2021)

\bibitem{Gianfagna2023} 
Gianfagna G. \textit{et al.}, MNRAS, \textbf{518}, 4238 (2023)

\bibitem{Capalbo2021}
V. Capalbo \textit{et al.}, MNRAS, \textbf{503}, 6155 (2021)

\bibitem{SZ}
R. A. Sunyaev, I. B. Zeldovich, ARA\&A, \textbf{18}, 537 (1980)

\bibitem{Cui2018}
W. Cui \textit{et al.}, MNRAS, \textbf{480}, 2898 (2018)

\bibitem{Haggar2020}
Haggar R. \textit{et al.}, MNRAS, \textbf{492}, 6074 (2020)

\bibitem{DeLuca2021}
F. De Luca \textit{et al.}, MNRAS, \textbf{504}, 5383 (2021)

\bibitem{Noll1976}
R. J. Noll, J. Opt. Soc. Am., \textbf{66}, 207 (1976)

\bibitem{Svechnikov2015}
M. Svechnikov \textit{et al.}, Opt. Express, \textbf{23}, 14677 (2015)

\bibitem{PSZ2}
Planck Collaboration XXVII, A\&A, \textbf{594}, A27 (2016)

\bibitem{Planck_n_counts}
Planck Collaboration XXIV, A\&A, \textbf{594}, A24 (2016)

\bibitem{Planck_y_maps}
Planck Collaboration XXII, A\&A, \textbf{594}, A22 (2016)

\bibitem{deAndres2022}
de Andres D. \textit{et al.}, Nat. Astron., \textbf{6}, 1325 (2022)

\bibitem{Hurier2013}
G. Hurier \textit{et al.}, A\&A, \textbf{558}, A118 (2013)

\bibitem{Remazeilles2011}
M. Remazeilles \textit{et al.}, MNRAS, \textbf{410}, 2481 (2011)

\bibitem{Planck_bias}
Planck Collaboration XX, A\&A, \textbf{571}, A20 (2014)


\end{thebibliography}
%
%

\end{document}